\input macros.sty
\documentclass{PoS}

\usepackage{graphicx,cite}

\title{Wilson fermions at fine lattice spacings: scale setting, pion
  form factors and $(g-2)_\mu$}

\ShortTitle{Wilson fermions at fine lattice spacings}

\author{Bastian B.~Brandt, Stefano Capitani, Dalibor
        Djukanovic, Georg von Hippel, Benjamin J\"ager,
        Bastian~Knippschild\\
        Institut f\"ur Kernphysik, University of Mainz, Becher Weg 45,
        55099 Mainz, Germany}

\author{Michele Della Morte, \speaker{Hartmut Wittig} \\
        Institut f\"ur Kernphysik and Helmholtz Institute Mainz,
        University of Mainz, Becher Weg 45, 55099 Mainz, Germany\\
        E-mail: \email{wittig@kph.uni-mainz.de}}

\author{Andreas~J\"uttner\\
        Physics Department, TH Unit, 1211 Geneva 23, Switzerland}

\abstract{
\vspace{-12cm}
          We present an update on our on-going project to compute hadronic
          observables for $\Nf=2$ flavours of $\rmO(a)$ improved
          Wilson fermions at small lattice spacings. The procedure to
          determine the lattice scale via the mass of the Omega baryon
          is described. Furthermore we present preliminary results for
          the pion form factor computed using twisted boundary
          conditions, and report on the implementation of a novel
          approach to determine the contribution of the hadronic
          vacuum polarisation to the anomalous magnetic moment of the
          muon.
\vspace{-17cm}{\begin{flushright} {\tt MKPH-T-10-18\\ HIM-2010-03 \\
          CERN-PH-TH-2010-228}\end{flushright}} 
}

\FullConference{The XXVIII International Symposium on Lattice Field Theory, Lattice2010\\
		June 14-19, 2010\\
		Villasimius, Italy}

\begin{document}

\section{Introduction \label{s1intro}}
Studies of structural properties of mesons and baryons, encoded in
their electromagnetic form factors, are the subject of many recent
lattice calculations. However, systematic uncertainties arising from
a variety of sources appear to be much larger than for simpler
quantities such as hadron masses and decay constants. As part of the
CLS project \footnote{{\tt
https://twiki.cern.ch/twiki/bin/view/CLS/WebHome}} a set of ensembles
is being generated, which will allow for a thorough investigation of
systematic effects, such as lattice artefacts, finite-volume effects,
and extrapolations in the light quark masses. The CLS project is
based on proven and conceptually simple technology: simulations are
performed using non-perturbatively $\rmO(a)$ improved Wilson quarks
and the Wilson plaquette action. In order to preserve the local
structure of the lattice action, the link variables remain
unsmeared. We employ the deflation-accelerated DD-HMC algorithm
\cite{DDHMC-defl} for $\Nf=2$ flavours of light quarks.

A potentially severe problem for any lattice simulation near the
continuum limit is the observed sharp rise in the autocorrelation time
of the topological charge\,\cite{ac:topol}. This important issue has
been addressed in several papers\,\cite{MLflow,SchSoVir10} and
contributions to this conference\,\cite{ML_lat10}. In spite of the
progress made in identifying the causes of this phenomenon, a rigorous
and effective solution is still elusive. Therefore we have restricted
the calculation of observables to a range of bare couplings for which
the problem of very long autocorrelation times in the topological
charge is not observed. A compilation of these ensembles is shown in
Table\,\ref{tab:params}. At our smallest value of the lattice spacing,
$a\approx0.05\,\fm$ (i.e. at $\beta=5.5$), where the problem is
expected to be most severe, we have checked explicitly that the
topological charge fluctuates around zero at a sizeable rate and
produces a reasonably symmetric distribution \cite{wittig_lat09}. We
can therefore take confidence that the composition of our ensembles is
not strongly biased and that our statistical errors are reliable.

\begin{table}[b]
\begin{center}
\begin{tabular}{ccccccc}
\noalign{\vskip0.3ex}
\hline\hline\noalign{\vskip0.3ex}
 $\beta$ &    $a[\fm]$    & lattice   &  $L [\fm]$ & \# masses   &
 $m_{\pi}L$ & Labels \\
\noalign{\vskip0.3ex}
\hline\noalign{\vskip0.3ex}
  5.20 & 0.08 & $64 \times 32^3 $ & 2.6 &  4 masses & 4.8 -- 9.0 &
  $\sf A1 - A4$ \\ 
\hline\noalign{\vskip0.3ex}
  5.30 & 0.07 & $48 \times 24^3 $ & 1.7 &  3 masses & 4.6 -- 7.9 &
  $\sf D1 - D3$ \\ 
  5.30 & 0.07 & $64 \times 32^3 $ & 2.2 &  3 masses & 4.7 -- 7.9 &
  $\sf E3 - E5$ \\ 
  5.30 & 0.07 & $96 \times 48^3 $ & 3.4 &  2 masses & 5.0, 4.2   &
  $\sf F6, F7$ \\ 
\hline \noalign{\vskip0.3ex}
  5.50 & 0.05 & $96 \times 48^3 $ & 2.5 &  3 masses & 5.3 -- 7.7 &
  $\sf N3 - N5$ \\ 
  5.50 & 0.05 & $128\times 64^3 $ & 3.4 &  1 mass   & 4.7 & $\sf O6$
  \\ \noalign{\vskip0.3ex}
\hline\hline
\end{tabular}
\caption{\label{tab:params} Simulation parameters and approximate
  values for the lattice scale and pion masses for those ensembles
  which show an acceptable tunnelling rate of the topological
  charge. The preliminary results presented in theses proceedings are
  based on the ensembles labelled ``N'', ``E'' and ``F6''.}
\end{center}
\end{table}

In this note we report on preliminary results for the pion
electromagnetic form factor.\footnote{Form factor calculations for
baryons are presented in another contribution to this
conference\,\cite{BKnipp_lat10}.} Another issue which has received a
lot of attention recently, is the hadronic vacuum polarisation
contribution to the muon's anomalous magnetic moment,
$a_\mu^{\rm{had}}$. Lattice calculations of this quantity involve the
determination of the vacuum polarisation amplitude, which, like
hadronic form factors, depends on a momentum variable. Therefore, we
discuss both the calculation of mesonic form factors and the
determination of $a_\mu^{\rm{had}}$ in these proceedings.

\section{Setting the scale \label{s2scale}}
\vspace{-0.2cm}
Studying the scaling properties of observables and expressing
dimensionful quantities in physical units requires the computation of
a reference quantity which sets the scale. The mass of the $\Omega$
baryon has emerged as a good candidate, since the $\Omega$ is stable
in QCD, and its valence sector consists entirely of strange quarks.
Moreover, ChPT studies \cite{Tiburzi:2005na} suggest a simple
functional form for chiral extrapolations in the sea quark mass.

\begin{figure}
\begin{center}
\vspace{-0.5cm}
\leavevmode
\includegraphics[height=5.6cm]{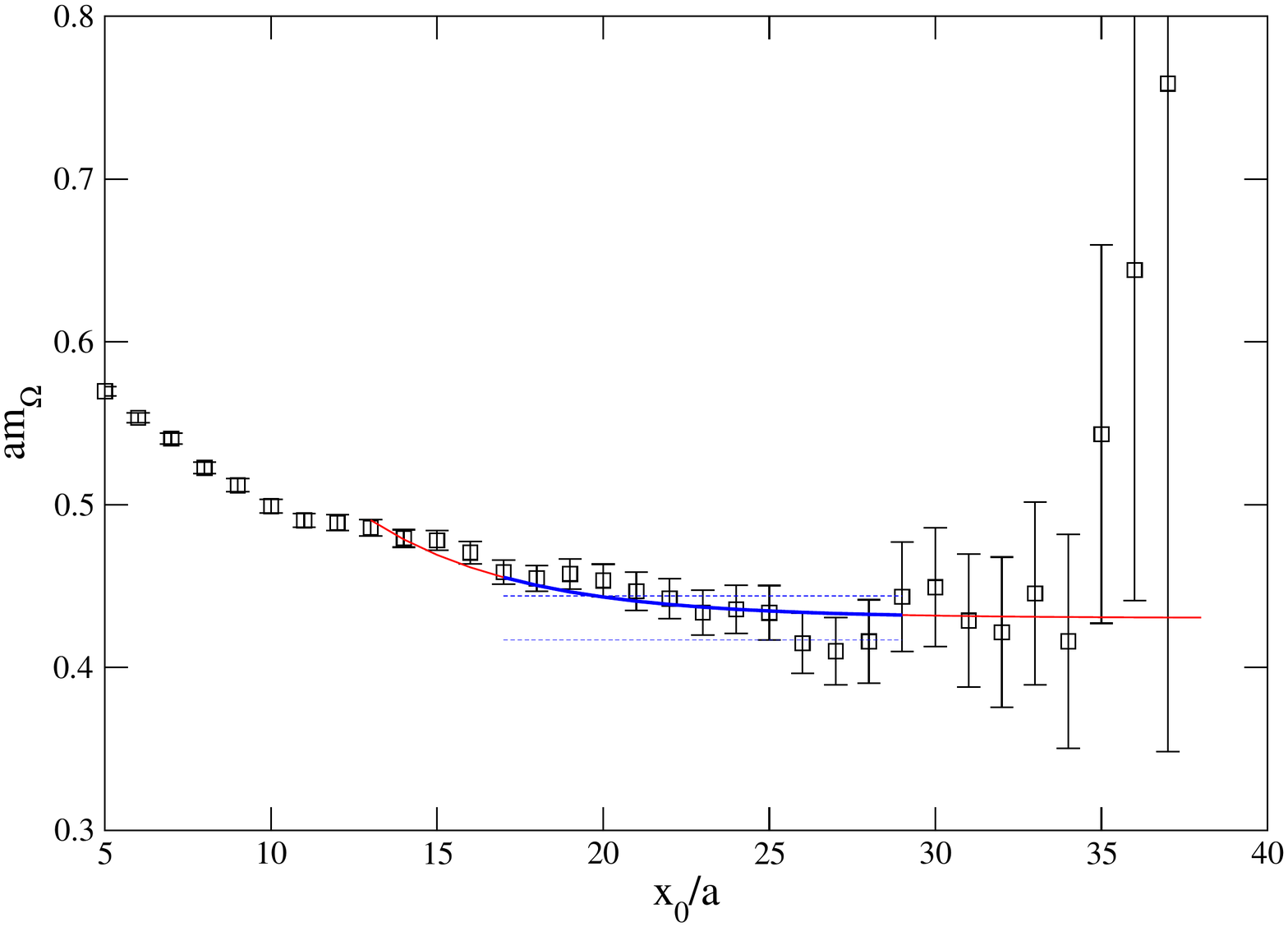}
\leavevmode\hspace{0.2cm}
\includegraphics[height=5.6cm]{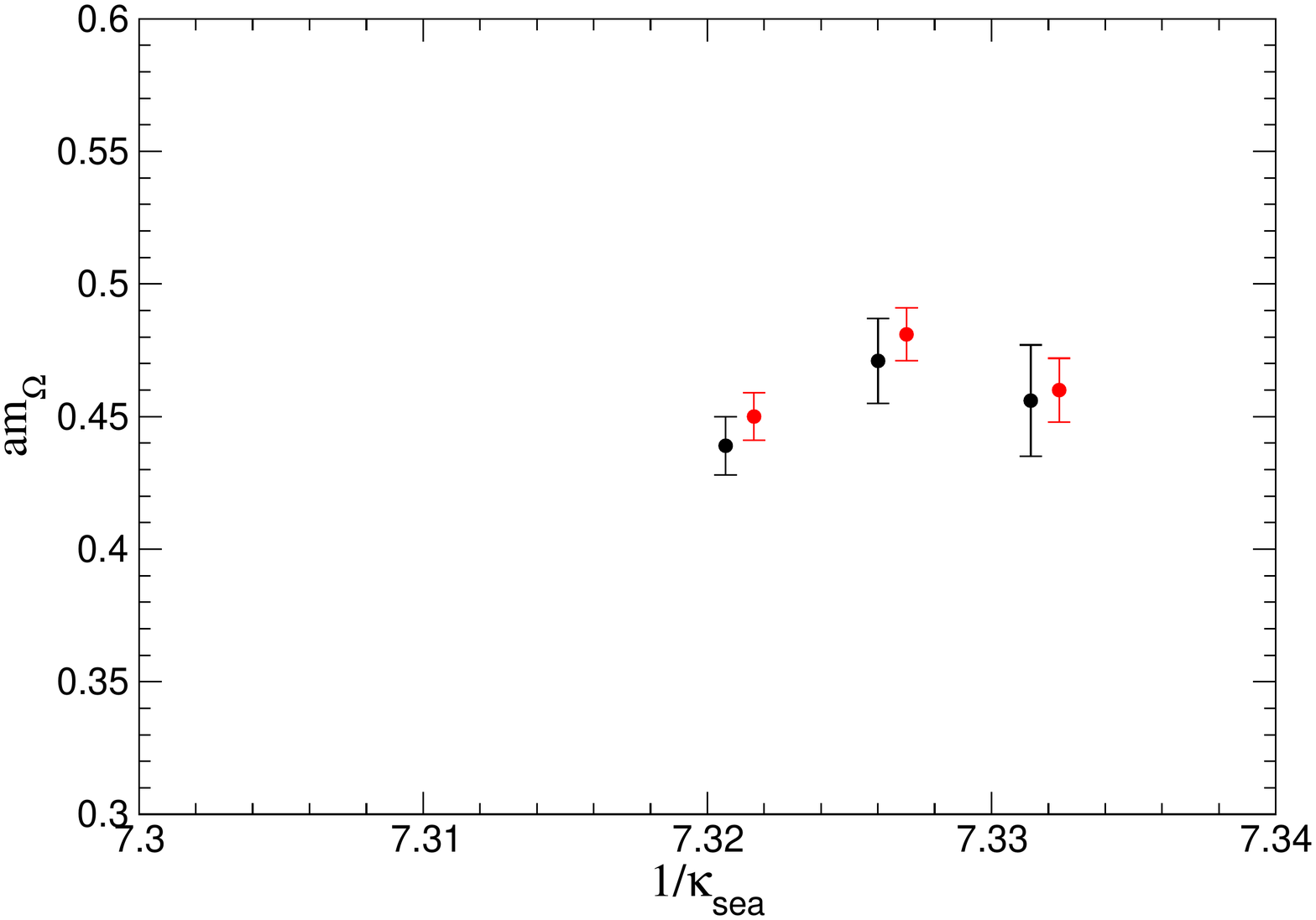}
\caption{\label{fig:Omega} {\bf Left:} the effective mass in the
  $\Omega$-channel for the ensemble~N5 ($m_\pi\approx410\,\MeV$); {\bf
  Right:} Sea quark mass dependence of $m_\Omega$ for datasets $\rm
  N3-N5$. Red and black symbols denote the results from correlated and
  uncorrelated fits, respectively.}
\vspace{-0.5cm}
\end{center}
\end{figure}

In order to determine the masses of pseudoscalar and vector mesons, as
well as octet and decuplet baryons, we used Jacobi-smeared sources
\cite{smear:Jacobi93}, supplemented by HYP-smeared \cite{smear:HYP01}
link variables. In the pseudoscalar channel long and stable plateaus
were observed, which allowed for a reliable determination of the mass
via single-cosh fits to the correlation functions. In the vector
channel and for baryons, the unambiguous identification of the ground
state turned out to be more difficult. We therefore employed the
procedure of \cite{Debbio2}, which is based on an {\it ansatz} that
includes contributions from the first excited state. In
Fig.\,\ref{fig:Omega} we show a typical effective mass plot in the
$\Omega$-channel together with the fit result.

In order to determine $m_\Omega$ an interpolation in the valence quark
mass must be performed. This requires knowledge of the hopping
parameter, $\kappa_s$, which corresponds to the bare strange quark
mass. We have determined $\kappa_s$ using the procedure described in
\cite{Debbio1,wittig_lat09}: Denoting the generic non-degenerate
pseudoscalar and vector mesons by $m_K$ and $m_K^\ast$, respectively,
we interpolate the mass ratio $(m_K/m_K^\ast)^2$ as a function of
$(am_K)^2$ to the experimental value $m_K/m_K^\ast=0.554$. This yields
the value of $\kappa_s$ at each value of the sea quark mass. In a
second step one interpolates $am_\Omega$ in the valence quark mass to
the value of $\kappa_s$. The resulting estimates of $am_\Omega$ as a
function of $1/\kappa_{\rm{sea}}$ are shown for $\beta=5.5$ in the
right panel of Fig.\,\ref{fig:Omega}. A mild dependence on the sea
quark mass is observed. At this stage we have not yet attempted a
chiral extrapolation in the sea quark mass to the physical value of
$m_\pi$. In order to set the scale at $\beta=5.5$ we therefore take
the result for $am_\Omega$ at the smallest sea quark, which yields
$a_{\Omega}=0.053(1)\,\fm$.

Our production runs for baryonic two-point functions on the other
ensembles are not yet completed. In order to compare form factors
computed at different lattice spacings consistently, we employ a
simple scaling relation, i.e.
\be
   \left. a_{\Omega}\right|_{\beta=5.3} =
   \left( \frac{a_{\rm{ref}}|_{\beta=5.3}}{a_{\rm{ref}}|_{\beta=5.5}}
   \right)
   \left. a_{\Omega}\right|_{\beta=5.5}.
\label{eq:scale}
\ee
Here the scale $a_{\rm{ref}}$ is defined via the kaon mass $am_K$,
evaluated at the (unphysical) reference point where $m_\pi/m_K=0.85$
\cite{Debbio1}, which can be reached without performing a chiral
extrapolation. At $\beta=5.3$ this yields
$a_{\rm{ref}}=0.0784(10)\,\fm$ \cite{Debbio1}, while at $\beta=5.5$
one obtains $a_{\rm{ref}}=0.0603(15)\,\fm$ \cite{wittig_lat09}. Our
result for $a_{\Omega}$ at $\beta=5.5$ and the scaling relation
\eq{eq:scale} imply
\be
   \left. a_{\Omega}\right|_{\beta=5.5}=0.053(1)\,\fm,\qquad
   \left. a_{\Omega}\right|_{\beta=5.3}=0.069(2)\,\fm.
\label{eq:afm}
\ee
%
%
At $\beta=5.5$ the value of $a_\Omega$ is smaller by more than 10\%
compared to $a_{\rm{ref}}$, which we used previously to convert our
pion masses on the ``N''-lattices into physical units (see Table\,2 of
ref.\,\cite{wittig_lat09}). Using the value in \eq{eq:afm} the pion
masses change to $m_\pi=600,\,510$ and $410\,\MeV$ on N3, N4 and N5,
respectively.

\section{The pion electromagnetic form factor \label{s3pionff}}

The electromagnetic form factor, defined by
\be
   \left\langle\pi^+(\pvec_f)|
   {\textstyle\frac{2}{3}}\ubar\gamma_\mu u
  -{\textstyle\frac{1}{3}}\dbar\gamma_\mu d
   |\pi^+(\pvec_i)\right\rangle = (p_f+p_i)_\mu\,f_\pi(q^2),
\ee
where $q=p_f-p_i$ is the momentum transfer, encodes the distribution
of electric charge inside the pion. Of particular interest is the
charge radius, $\langle r^2_\pi\rangle$, which is derived from the
pion form factor at vanishing momentum transfer, i.e.
\be
       f_\pi(q^2)=1-\frac{1}{6}\langle r_\pi^2\rangle
       q^2+\rmO(q^4) \quad\Rightarrow\quad
       \langle r^2_\pi\rangle = 6\left.
       \frac{\rmd f_\pi(q^2)}{\rmd q^2}\right|_{q^2=0}.
\ee
Lattice calculations of mesonic matrix elements are technically
simpler than the corresponding quantities for the
nucleon. Furthermore, the pion electromagnetic form factor receives no
contributions from quark-disconnected diagrams. This opens the
possibility to perform a precision test of lattice QCD, by comparing
lattice estimates for $\langle r^2_\pi\rangle$ to the experimentally
determined value.
However, owing to the finite spatial volume the accessible range of
momentum transfers $Q^2=-q^2$ is severely constrained, which presents
a major obstacle for precise lattice determinations of $\langle
r^2_\pi\rangle$. We have therefore employed flavour-twisted boundary
conditions\,\cite{twbc} in the valence sector, such that the momentum
transfer satisfies\,\cite{FFtwbc:UKQCD07}
\be
    -Q^2\equiv q^2=(p_f-p_i)^2 =
   \Big(E_\pi(\vec{p}_f)-E_\pi(\vec{p}_i)\Big)^2
  -\Big[\Big(\vec{p}_f+\frac{\vec\theta_f}{L}\Big)
  -\Big(\vec{p}_i+\frac{\vec\theta_i}{L}\Big) \Big]^2.
\ee
We tuned the twist angles $\vec\theta_i,\,\vec\theta_f$ so as to
achieve a particularly fine momentum resolution near $Q^2=0$. In order
to improve the statistical signal we used stochastic $Z_2\times Z_2$
sources in the calculation of quark propagators. As in
ref.\,\cite{FFtwbc:UKQCD08} the pion form factor was extracted from a
suitable ratio of correlators in which the renormalisation factor of
the local vector current, $\zv$, drops out.

\begin{figure}
\begin{center}
\leavevmode
\includegraphics[height=5.6cm]{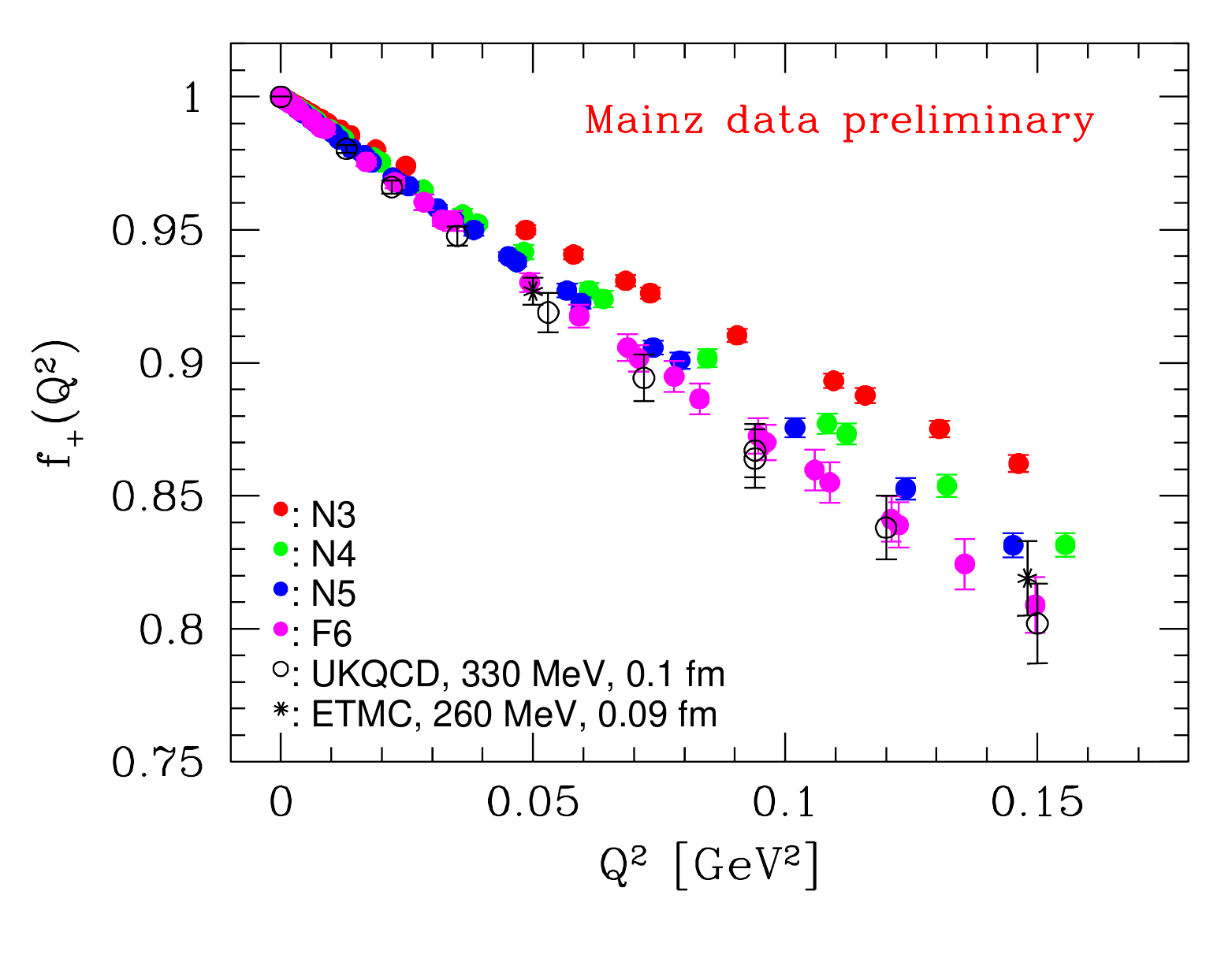}
\leavevmode\hspace{0.4cm}
\includegraphics[height=5.5cm]{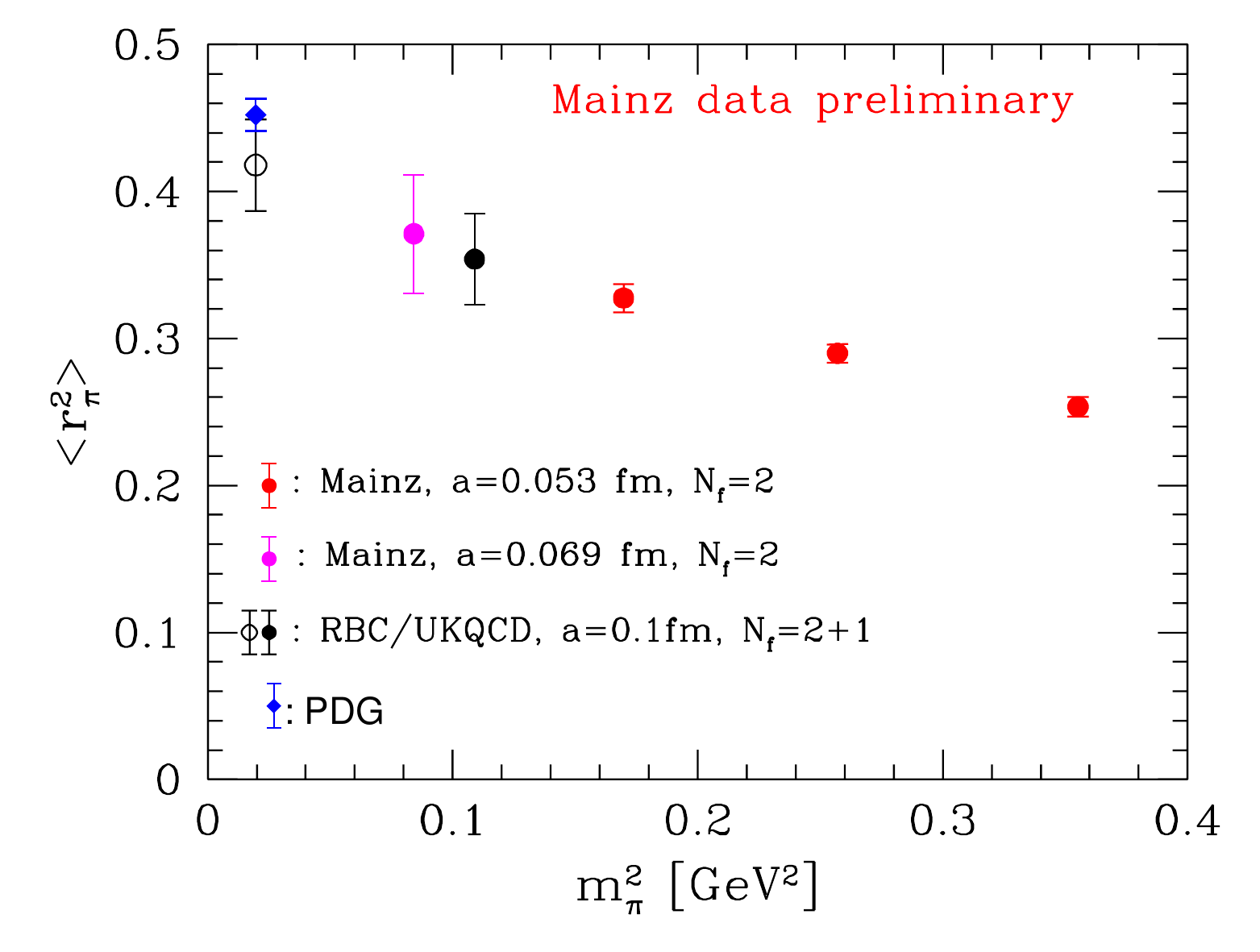}
\caption{\label{fig:pionff} {\bf Left:} pion form factor computed for
  a range of pion masses compared to the results of
  \cite{FFtwbc:UKQCD08,FFtwbc:ETM08}; {\bf Right:} the charge radius
  as a function of the pion mass, compared with the results of the
  $2+1$-flavour simulation from\,\cite{FFtwbc:UKQCD08}.}
\vspace{-0.5cm}
\end{center}
\end{figure}

In Fig.\,\ref{fig:pionff} we show our results for the pion
electromagnetic form factor as a function of the squared Euclidean
momentum transfer for the data sets N3, N4, N5 and~F6. The
corresponding pion masses in $\MeV$ are 600, 510, 410 and~290,
respectively. Twisted boundary conditions were also employed in
refs.\,\cite{FFtwbc:UKQCD08,FFtwbc:ETM08}, which are shown together
with our data. The much larger number of data points near $Q^2=0$
allows us to perform a very accurate and largely model-independent
determination of the pion charge radius, by computing the linear slope
of $f_\pi(Q^2)$ in the vicinity of $Q^2=0$. In this way our estimate
of the pion charge radius does not rely on a particular {\it ansatz},
such as vector dominance. As can be seen from the right panel in
Fig.\,\ref{fig:pionff} we obtain statistically very precise results
for $\langle r_\pi^2\rangle$, in particular on the ensembles at
$\beta=5.5$. Statistics at our lightest quark mass ($\beta=5.3$) will
be increased. Our preliminary results for $\langle r_\pi^2\rangle$ are
consistent with the determination of ref.\,\cite{FFtwbc:UKQCD08},
which uses $\Nf=2+1$ flavours of domain wall quarks, and the trend in
our lattice data compares favourably with the experimentally
determined value.

\section{Hadronic vacuum polarisation contribution to $(g-2)_\mu$
  \label{s4vacpol}} 

The anomalous magnetic moment of the muon,
$a_\mu\equiv\half(g-2)_\mu$, is among the most precisely measured
quantities. Assuming the validity of the Standard Model, the
experimental value for $a_\mu$ differs from the theoretical prediction
at the level of 3.2~standard deviations \cite{Jegerlehner:2009ry}. An
important ingredient is the leading hadronic contribution to vacuum
polarisation, $a_\mu^{\rm{had}}$. This quantity is normally determined
via a semi-phenomenological approach based on the evaluation of a
dispersion integral containing experimentally measured hadronic cross
sections. Given the importance of $a_\mu$ for new physics searches, a
first-principles determination of $a_\mu^{\rm{had}}$ is highly
desirable. Following refs.\,\cite{amu:EdeRaf93,amu:Blum02},
$a_\mu^{\rm{had}}$ can be computed on the lattice via the convolution
integral
\be
  a_\mu^{\rm had} = 4\pi^2\left(\frac{\alpha}{\pi}\right)^2
  \int_0^{\infty} {\rmd}Q^2\,f(Q^2)\big\{ \Pi(Q^2)-\Pi(0) \big\},
\label{eq:amuhad}
\ee
where $\Pi(Q^2)$ is related to the vacuum polarisation tensor
$\Pi_{\mu\nu}$, i.e.
\be
  \Pi_{\mu\nu}(Q) = \int\rmd^4{x}\,
  \rme^{iq\cdot(x-y)}\,\left\langle J_\mu(x) J_\nu(y) \right\rangle
  \equiv (q_\mu q_\nu -g_{\mu\nu}q^2) \Pi(q^2).
\ee
Here $J_\mu$ denotes the flavour-singlet vector current. A well-known
problem is that the convolution function $f(Q^2)$ is strongly peaked
for momenta near $m_\mu$, which is an order of magnitude smaller than
what can conventionally be realised on current lattice sizes. Twisted
boundary conditions cannot be used in a straightforward manner, since
the current-current correlator contains quark-disconnected diagrams,
for which the effect of the twist cancels.

To tackle this problem we have implemented the strategy outlined in
\cite{amu:morte_juett}: The key observation is that the continuum
limit exists for the individual quark-disconnected and connected
contributions to $\Pi$. Moreover, the connected part can be
re-interpreted in terms of flavour-non-singlet correlators, which are
easily computed using twisted boundary conditions in the standard
fashion. ChPT at NLO also predicts that the contribution from
disconnected diagrams is suppressed by a factor $-1/10$ relative to
the connected ones\,\cite{amu:morte_juett}.

\begin{figure}
\begin{center}
\leavevmode
\includegraphics[height=5.1cm]{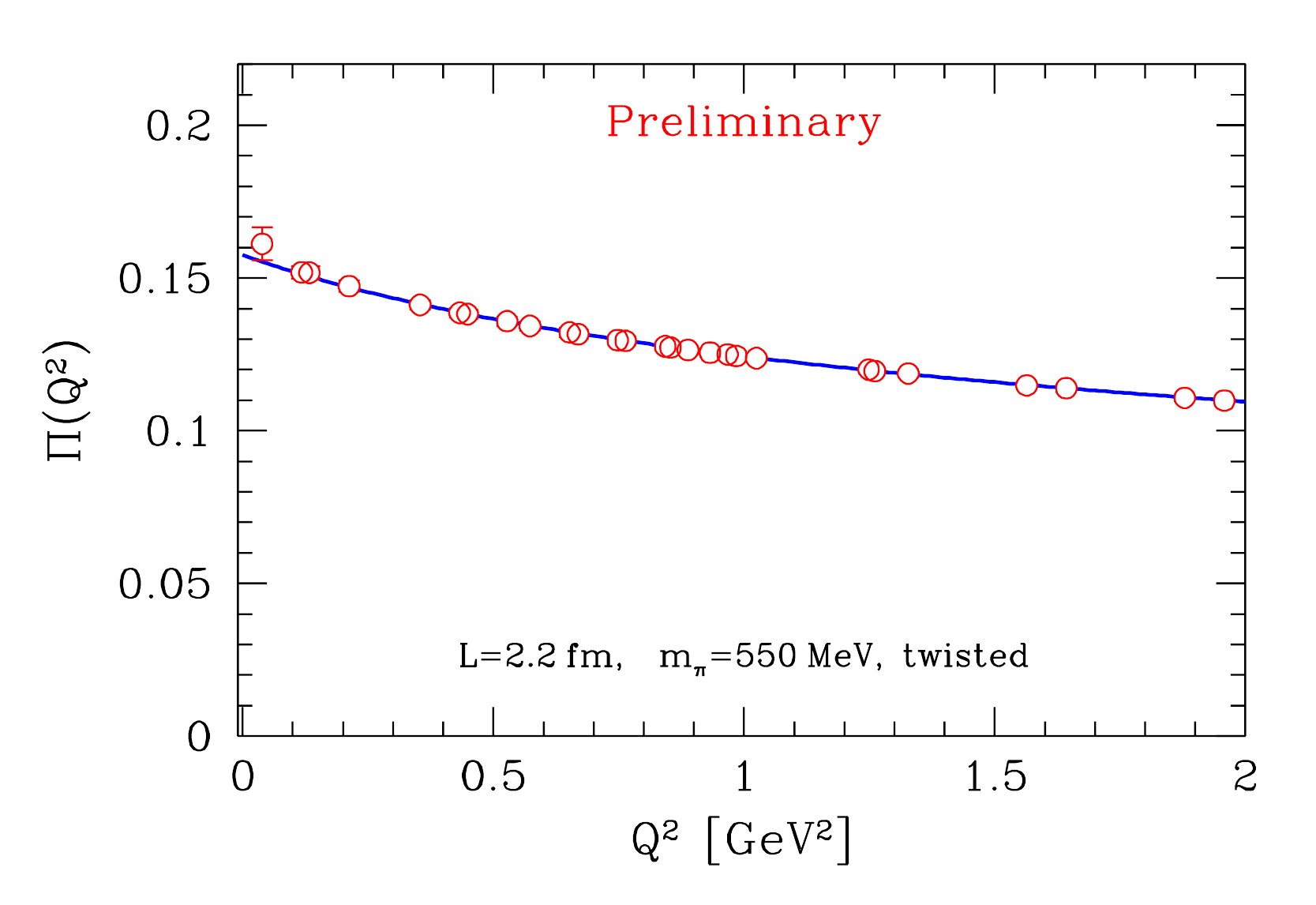}
\leavevmode\hspace{0.0cm}
\includegraphics[height=5.1cm]{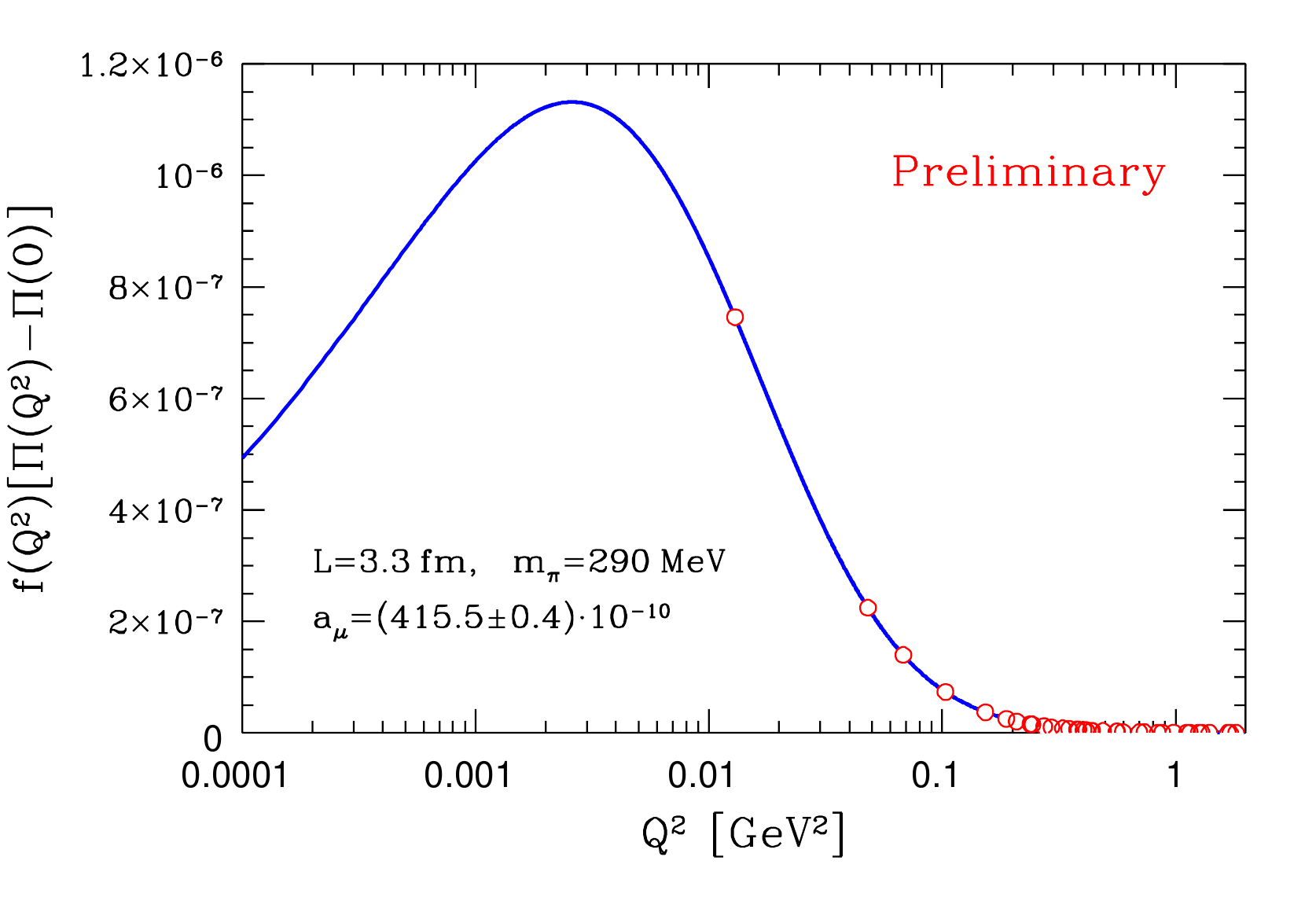}
\caption{\label{fig:amu} {\bf Left:} the vacuum polarisation amplitude
  computed at $\beta=5.3$, $L\simeq2.2\,\fm$, $m_\pi\approx550\,\MeV$,
  using twisted boundary conditions. The solid curve is a Pad\'e fit
  to the data; {\bf Right:} the integrand of the convolution integral
  on the ``F6'' lattice. The area underneath the solid curve yields
  the value of $a_\mu^{\rm{had}}$.} 
\vspace{-0.5cm}
\end{center}
\end{figure}

Preliminary results for the vacuum polarisation amplitude extracted
from the connected correlator in the two-flavour case are shown in
Fig.\,\ref{fig:amu}. Twisted boundary conditions produce a much larger
density of $Q^2$ values, which greatly stabilises the extrapolation to
$Q^2=0$ and the extraction of $\Pi(0)$, which enters
\eq{eq:amuhad}. Furthermore, owing to the larger reach in $Q^2$ the
integrand of the convolution integral is much more tightly constrained
(c.f. Fig.\,\ref{fig:amu}, right panel). Our lattice estimates for
$a_\mu^{\rm{had}}$ computed for $\Nf=2$ and also in the theory
containing a quenched strange quark will appear
elsewhere\,\cite{in_prep}. Eventually we will also include
quark-disconnected diagrams, computed for the usual Fourier-momenta
only, in order to verify that their contribution is indeed suppressed.

\vspace{-0.2cm}
\section{Summary and outlook \label{s5summ}} 
\vspace{-0.2cm}
The ensembles generated as part of the CLS project allow for a
comprehensive investigation of systematic effects for a variety of
hadronic observables. Our fine lattice resolution will enable reliable
continuum extrapolations and will prove useful to explore hadronic
form factors at large momentum transfers. Twisted boundary conditions
are indispensable for an accurate determination of the pion form
factor. Their use can also be extended to lattice determinations of
$a_\mu^{\rm{had}}$, for which a straightforward application of this
technique is not obvious.
\medskip
\par\noindent{\bf Acknowledgments:} We thank our colleagues within the
CLS project for sharing gauge ensembles. Calculations of correlation
functions were performed on the dedicated QCD platform ``Wilson'' at
the Institute for Nuclear Physics, University of Mainz. This work is
supported by DFG (SFB443), GSI, and the Research Center~EMG funded by
{\sl Forschungsinitiative Rheinland-Pfalz}.

\providecommand{\href}[2]{#2}\begingroup\raggedright\endgroup


\end{document}